\documentclass[11pt,preprint]{aastex}
\usepackage{amsmath}
\newcommand{\beq}{\begin{equation}}
\newcommand{\eeq}{\end{equation}}
\def\gtsim {\lower .1ex\hbox{\rlap{\raise .6ex\hbox{\hskip .3ex
        {\ifmmode{\scriptscriptstyle >}\else
                {$\scriptscriptstyle >$}\fi}}}
        \kern -.4ex{\ifmmode{\scriptscriptstyle \sim}\else
                {$\scriptscriptstyle\sim$}\fi}}}

\shorttitle{Earth mass micro-haloes}
\shortauthors{Moore et al.}

\begin{document}

\title{On the survival and disruption of Earth mass CDM micro-haloes}

\author{Ben Moore\altaffilmark{1}, 
  Juerg Diemand\altaffilmark{1,2},
 Joachim Stadel\altaffilmark{1}
 \& Thomas Quinn\altaffilmark{1,3}\\
 }

\altaffiltext{1}{Institute for Theoretical Physics, University of Z\"urich, 
CH-8057 Z\"urich, Switzerland} 
\altaffiltext{2}{Dept. of Astronomy and Astrophysics,
       University of California,
       Santa Cruz, CA 95064, USA}
\altaffiltext{3}{Dept. of Astronomy,
       University of Washington,
       Seattle, Box 351580, USA}

\begin{abstract}

Neutralino dark matter leads to the formation of numerous earth mass
dark matter haloes at redshifts $z\approx 60$
\citep{Diemand2005}. These abundant CDM micro-haloes have cuspy
density profiles that can easily withstand the Galactic tidal field at
the solar radius.  Zhao, Taylor, Silk \& Hooper (astro-ph/0502049)
concluded that ``...the majority of dark matter substructures with
masses $\sim 10^{-6}M_\odot$ will be tidally disrupted due to
interactions with stars in the Galactic halo''.  However these authors
assumed a halo density of stars that is at least an order of magnitude
higher than observed. We show that the appropriate application of the
impulse approximation is to the regime of multiple encounters, not
single disruptive events as adopted by Zhao et al., which leads to a
survival time of several Hubble times.  Therefore we do not expect the
tidal heating by Galactic stars to affect the abundance of
micro-haloes.  Disk crossing will cause some mass loss but the central
cores are likely to survive and could be detected as gamma-ray sources
with proper motions of several arc minutes per year.

\end{abstract}

\section{Discussion}
\label{section:introduction}

The rate of energy transfer by tidal heating by a class of objects depends on 
several factors including their mean density ${n_*}$ and mean mass ${m_*}$.
Zhao et al. assumed $n_*m_*=0.1$ for the Galactic disk and $n_*m_*=0.001$ for 
halo stars. Observations suggest that more appropriate numbers would be $0.05$ and 
$<0.0001$ for disk and halo stars respectively, \citep{MB1981, Flynn1996, Gould1998, Ivezic2000, Spagna2004}. 
Here the halo density is inferred from star counts, proper 
motion surveys, deep HST images and Sloan RR Lyrae counts.
In their revised comment (version \#4) Zhao et al. replaced the ``halo'' contribution
with a ``bulge'' component. 
However this does not change the observational evidence that the stellar density
high above the local Galactic disk is very low. Furthermore
extrapolating the Galactic bulge to the solar radius is premature since
recent observations suggest that the bulge is really part of the Galactic bar \citep{Fr1998, Pi2004},
truncated at 4 kpc that is possibly undergoing a 
complex buckling instability. In which case it is unlikely that the bulge
contributes to the halo density at the solar radius.
These estimates of the halo density are all at least an order of magnitude lower 
than assumed
by Zhao et al., which reduces the heating rate by non-disk stars to a minor level.

For Earth mass micro-halos with a half mass radius $r_{1/2}=0.005$ pc 
and $m_{halo}=10^{-6}M_\odot$ we find a critical impact parameter with a star, such
that its internal energy changes by of order itself, to be $b_{min}=0.012$ pc.
The disk surface density is observed to be $\approx 46 M_\odot {\rm pc}^{-2}$ 
\citep{K1989}, therefore
the probability of a disruptive collision per disk crossing is less than $0.02$. 
This neglects the clustering of stars in the disk and is a crude estimate of survival
probability but demonstrates that most of the haloes are not
in the single encounter disruption regime.
In Diemand et al. (2005) we estimated the disruption timescale by
integrating the impulsive heating $\Delta E \propto G^2r_{1/2}^2m_*^2m_{halo}/(b^4v^2)$ over all encounters
and comparing to the binding energy of CDM haloes $E_{bind}\propto Gm_{halo}^2/r_{1/2}$ i.e. Binney \& Tremaine (1989). 
With appropriate factors for the internal structure of these haloes we found a disruption 
timescale of $t_{dis}\approx 30$ Gyrs, assuming 
that the micro-haloes spend one percent of their time in the disk.

As mentioned in Diemand et al. (2005) we expect encounters with stars and molecular clouds 
to lead to some disruption and mass loss but that most such structures will stay intact.
The impulse approximation is an approximation, as are semi-analytic models for such 
mass loss -- tidal heating 
rarely leads to complete
disruption and the high density central cores of orbiting systems can easily 
remain intact. Numerical simulations are vital to understand these complex dynamical processes.
A more detailed study supported
by numerical simulations in the spirit of Moore (1993) 
by the authors is in preparation.
The most important factor in the survival statistics of micro-haloes is how many survive
similar mass mergers as a Galactic mass halo is built up. 
However, even if just a few percent survive the hierarchical growth, many micro-haloes
will lie within one parsec from the sun. Their dense cuspy cores will 
be sources of gamma-ray emission which could be uniquely distinguished
by their high proper motions of order minutes of arc per year.

\bibliography{ms}

\begin{thebibliography}{39}
\expandafter\ifx\csname natexlab\endcsname\relax\def\natexlab#1{#1}\fi

\parskip=-4pt

\bibitem[Binney \& Tremaine(1987)]{BHT} Binney, J., \& 
Tremaine, S.\ 1987, Princeton, NJ, Princeton University Press, 1987

\bibitem[Diemand et al.(2005)]{Diemand2005}
Diemand, J., Moore, B. \& Stadel. 2005, Nature, 433, 389-391

\bibitem[Flynn et al.(1996)]{Flynn1996} Flynn, C., Gould, A., \& 
Bahcall, J.~N.\ 1996, \apjl, 466, L55 

\bibitem[Freudenreich(1998)]{Fr1998} Freudenreich, H.~T.\ 
1998, \apj, 492, 495 

\bibitem[Kuijken \& Gilmore(1989)]{K1989} Kuijken, K., \& 
Gilmore, G.\ 1989, \mnras, 239, 605 

\bibitem[Gould et al.(1998)]{Gould1998} Gould, A., Flynn, C., \& 
Bahcall, J.~N.\ 1998, \apj, 503, 798 

\bibitem[Ivezi{\' c} et al.(2000)]{Ivezic2000} Ivezi{\' c}, {\v 
Z}., et al.\ 2000, \aj, 120, 963 

\bibitem[Mihalas \& Binney(1981)]{MB1981} Mihalas, D., \& 
Binney, J.\ 1981, San Francisco, CA, W.~H.~Freeman and Co., 1981.~608  

\bibitem[Moore(1993)]{Moore1993} Moore, B.\ 1993, \apjl, 413, L93 

\bibitem[Picaud \& Robin(2004)]{Pi2004} Picaud, S., \& Robin, 
A.~C.\ 2004, \aap, 428, 891 

\bibitem[Spagna et al.(2004)]{Spagna2004} Spagna, A., Carollo, D., 
Lattanzi, M.~G., \& Bucciarelli, B.\ 2004, \aap, 428, 451 


\end{thebibliography}

\clearpage

\end{document}